\documentclass[preprint2]{aastex}

\usepackage{psfig}
\usepackage{amsmath}

\newcommand{\chandra}{\textit{Chandra}}
\newcommand{\chandralong}{\textit{Chandra X-ray Observatory}}
\newcommand{\xmmlong}{\textit{XMM-Newton}}
\newcommand{\xmm}{\textit{XMM}}

\newcommand{\nh}{\mbox{$N_{\rm H}$}}

\newcommand{\rinfty}{\mbox{$R_{\infty}$}}
\newcommand{\rns}{\mbox{$R_{\rm NS}$}}
\newcommand{\mns}{\mbox{$M_{\rm NS}$}}

\newcommand{\chisq}{\mbox{$\chi^2$}}
\newcommand{\chisqnu}{\mbox{$\chi^2_\nu$}}

\newcommand{\Msun}{\mbox{$M_\odot$}}

\newcommand{\CstR}{Cst\rns}

\newcommand{\mr}{\mbox{\mns--\rns}}
\newcommand{\MofR}{\mbox{\mns(\rns)}}

\newcommand{\xray}{\mbox{X-ray}}

\newcommand{\simlt}{\mathrel{\hbox{\rlap{\hbox{\lower4pt\hbox{$\sim$}}}\hbox{$<$}}}}
\newcommand{\simgt}{\mathrel{\hbox{\rlap{\hbox{\lower4pt\hbox{$\sim$}}}\hbox{$>$}}}}
\newcommand{\approxgt}{\mbox{$\,^{>}\hspace{-0.24cm}_{\sim}\,$}}
\newcommand{\approxlt}{\mbox{$\,^{<}\hspace{-0.24cm}_{\sim}\,$}}
\newcommand{\ee}[1]{\mbox{$10^{#1}$}}

\newcommand{\ud}[2]{\mbox{$^{+ #1}_{- #2}$}}
\newcommand{\ppm}{\mbox{$\pm$}}

\newcommand{\unit}[1]{\mbox{$\rm\,#1$}}

\def\arcdeg{\hbox{$^\circ$}}

\newcommand{\msun}{\mbox{$\,M_\odot$}}

\newcommand{\km}{\hbox{$\,{\rm km}$}}

\newcommand{\MeV}{\mbox{$\,{\rm MeV}$}}

\newcommand{\ksec}{\mbox{$\,{\rm ks}$}}

\newcommand{\kpc}{\mbox{$\,{\rm kpc}$}}

\newcommand{\percmcube}{\mbox{$\,{\rm cm^{-3}}$}}

\newcommand{\cgsdensity}{\mbox{$\,{\rm g\percmcube}$}} 
\newcommand{\cgsaccel}{\mbox{$\,{\rm cm\,s^{-2}}$}}

\def\OmCen{\mbox{$\omega$\,Cen}}

\begin{document}
\title{Rejecting proposed dense-matter equations of state with
  quiescent low-mass X-ray binaries}

\author{Sebastien Guillot and Robert E. Rutledge } 

\affil{Department of Physics, McGill University,\\ 3600 rue
  University, Montreal, QC, Canada, H3A-2T8}

\email{guillots@physics.mcgill.ca, rutledge@physics.mcgill.ca}

\slugcomment{Draft - \today}
\shorttitle{Excluding dEOSs}

\begin{abstract}
Neutrons stars are unique laboratories for discriminating between the
various proposed equations of state of matter at and above nuclear
density. One sub-class of neutron stars -- those inside quiescent
low-mass \xray\ binaries (qLMXBs) -- produce a thermal surface
emission from which the neutron star radius (\rns) can be measured,
using the widely accepted observational scenario for qLMXBs, assuming
unmagnetized H atmospheres. In a combined spectral analysis, this work
first reproduces a previously published measurement of the \rns,
assumed to be the same for all neutron stars, using a slightly
expanded data set.  The radius measured is $\rns=9.4\pm1.2\km$. On the
basis of spectral analysis alone, this measured value is not affected
by imposing an assumption of causality in the core.  However, the
assumptions underlying this \rns\ measurement would be falsified by
the observation of any neutron star with a mass $>2.6\msun$, since
radii $<11\km$ would be rejected if causality is assumed, which would
exclude most of the \rns\ parameter space obtained in this analysis.
Finally, this work directly tests a selection of dense matter
equations of states: WFF1, AP4, MPA1, PAL1, MS0, and three versions of
equations of state produced through chiral effective theory.  Two of
those, MSO and PAL1, are rejected at the 99\% confidence level,
accounting for all quantifiable uncertainties, while the other cannot
be excluded at $>$99\% certainty.
\vspace{1.0cm}
\end{abstract}

\maketitle
\section{Introduction}

More than eighty years after the discovery of the neutron, much
uncertainty surrounds theoretical predictions related to strong-force
physics.  One prediction, the dense matter equation of state (dEOS
hereafter) relates the pressure and density of cold matter at and
above nuclear density $\rho_{0}\approxgt \ee{14}\cgsdensity$, such as
that found in the core of neutron stars (NSs).  The dEOS is still
actively debated among nuclear physicists \citep{lattimer07}, with
dEOS theories proposed based upon a variety of uncertain physical
processes and calculational methods
\citep{muther87,wiringa88,prakash88,muller96,akmal97,hebeler13}.
Because knowledge of the masses and radii of NSs permit discrimination
between dEOSs, these compact objects have been extensively observed
\citep[e.g.,][G13 hereafter]{lindblom92,ozel09b,guillot13}.

One sub-class of NSs -- those inside quiescent low-mass
\xray\ binaries (qLMXBs) -- produce thermal surface emission from
which constraints on the NS mass \mns\ and radius \rns\ can be
extracted \citep{brown98,rutledge99}.  In this widely invoked model,
the luminosity originates from energy deposited in the NS crust during
active accretion ($E\sim1.9\MeV$ per accreted nucleon
\citealt{haensel08}), by a series of electron captures, neutron
emissions and pycnonuclear reactions.  Once an active accretion
episode ends, this heat, mostly absorbed into the relatively cool
core, is then conducted to the surface on a core-cooling timescale,
before being radiated through the NS atmosphere \citep{brown98}.

Hydrogen atmosphere models for these NSs
\citep{rajagopal96,zavlin96,mcclintock04,heinke06a,haakonsen12}
consider the full radiative transfer through an accreted hydrogen
atmosphere of an unmagnetized NS, including the gravitational redshift
$1+z= \left(1-\frac{2G M_{\rm NS}}{R_{\rm NS} c^{2}}\right)^{-1/2}$
caused by the high surface gravity $\sim 10^{13}-10^{15}\cgsaccel$.
These models describe well the emergent spectra of qLMXBs and the
measured emission areas correspond to the $\sim$10\km\ radii expected
from NSs. This scenario has been found to be consistent with behavior
in multiple historical transient LMXBs in their quiescent phase
\citep[e.g. ][]{rutledge99,
  rutledge00,campana00a,intzand01,rutledge01a, rutledge01b,
  wijnands02,tomsick04, cackett06, jonker07a,cornelisse07,
  fridriksson10,degenaar14}.

While the higher signal-to-noise ratio (S/N) \xray\ spectra of qLMXBs
in the field of the galaxy (e.g., Cen~X-4, Aql~X-1) can potentially
place useful constraints on the dEOS from the measurement of the
projected radius $\rinfty=\rns \left(1+z\right)$, the 20--50\%
distance uncertainty to these sources directly contributes to a
20--50\% uncertainty of the \rinfty\ measurements.  Focusing on
qLMXBs located inside globular clusters (GCs), for which distances are
known to $\sim$5--10\% accuracy, decreases the associated uncertainty,
and so targets identified as qLMXBs in GCs are exclusively used for
this analysis.

Because of their relatively low-S/N -- but more importantly because of
the strong covariance between \mns\ and \rns\ in this spectral
interpretation -- the \xray\ spectra of qLMXBs inside GCs have
individually been hitherto unable to place stringent constraints on
the dEOS \citep{webb07,heinke06a}.  Combining these \xray\ spectra in
a single, simultaneous analysis increases the S/N; more significantly,
comparing an ensemble of \mr\ constraints directly to a single
proposed \MofR\ relation resulting from a theoretical dEOS produces
strong covariance between photon spectral model parameters in
different sources.

This results in a stronger confrontation between theory and data than
obtained when treating sources independently.  This was done in a
previous work to measure the radius of NSs, $\rns=9.1\ud{1.3}{1.5}\km$
(90\% confidence, G13), using five qLMXBs located in GCs, when it is
assumed that these NSs have a quasi-constant \rns\ (\CstR, hereafter),
i.e., the same value of \rns\ to within measurement uncertainties.
This assumption arises from the observational evidence
(\mns\ measurements of two $\sim2\msun$ pulsars,
\citealt{demorest10,antoniadis13}) that dense nuclear matter is best
described by dEOSs characterized by a quasi-constant \rns\ for
astronomically relevant masses ($\mns>0.5\msun$).

Significantly, an analysis \citep{steiner13} employing qLMXB spectral
parameters \citep{heinke06a,webb07,guillot11a} separately, and in
combination with photospheric radius expansion analysis spectral
parameters \citep{ozel09a,guver10a,guver10b,ozel12b} derived a
detailed formulation for the dEOS.  By assuming the dEOS would be
consistent with producing a measured $\sim$2\msun\ NS
\citep{demorest10}, it was found that the preferred NS
\mr\ relationship produces quasi-constant NS radii, demonstrating
consistency with this assumption \citep{steiner13}.  Furthermore, this
work showed that the radius of a 1.4\msun\ NS is contained within
11.23\km\ and 12.49\km\ (95\% confidence region), which, according to
that work, rules out $\sim$1/3 of Skyrme models as well as covariant
field-theoretical dEOS models, although without stating the quantified
probability with which these dEOS models are ruled out.

Here, widely used observational analyses are employed, but with a new
statistical analysis combining the data from six qLMXBs, comparing
them directly to \MofR\ relationships resulting from proposed dEOSs.
This method permits rejecting dEOSs with a quantifiable degree of
certainty.  Section~\ref{sec:analysis} summarizes the spectral
analysis, also available in greater detail in a previous work (G13),
and presents the additional data added to this work.
Section~\ref{sec:cstR} reproduces the results of the previous work
with the additional data, and adding the assumption of causality.
Section~\ref{sec:dEOSs} describes the analysis confronting a selection
of dEOS models to the spectral \xray\ data.  Finally,
Section~\ref{sec:ccl} discusses these results.

\begin{deluxetable}{lccccc}
  \tablecaption{\label{tab:constantR} Radius Measurements of Neutron Stars}
  \tablewidth{0cm}
  \tabletypesize{\scriptsize}
  \tablecolumns{6}
  \tablehead{
    \colhead{Description of } & \colhead{\rns\ (\km)} & \colhead{\rns\ (\km)} & \colhead{\chisq / d.o.f.} & \colhead{Null Hypothesis} & \colhead{Line style in} \\
    \colhead{Simulation}      & \colhead{90\% confidence}   & \colhead{99\% confidence}   & \colhead{}                  & \colhead{Probability}     & \colhead{Figure~\ref{fig:results} ({\it top})}
  }
  \startdata 
  Causality assumption    & 9.5\ud{1.2}{1.2} & 9.5\ud{1.9}{1.8} & 516 / 476 & 0.10 & Solid\\
  No causality assumption & 9.4\ud{1.2}{1.2} & 9.4\ud{1.9}{1.8} & 516 / 476 & 0.10 & Dashed\\
  \cite{guillot13}        & 9.1\ud{1.3}{1.5} & 9.1\ud{2.0}{2.2} & 613 / 628 & 0.64 & Dotted
  \enddata

  \tablecomments{No significant variation of the measured \rns\ is
    observed, when additional data are added (for qLMXBs in M30 and
    \OmCen) to the spectra used in the previous work (G13). Assuming
    causality does not change \rns\ by more that 1\%.  For the
    analysis with the causality assumption, MCMC samples violating the
    condition $\rns < 2.83 G\mns / c^{2}$ are excluded.}
\end{deluxetable}

\section{Simultaneous spectral fitting of qLMXBs with H-atmosphere models}
\label{sec:analysis}

The simultaneous spectral analysis of the qLMXBs is performed using
the Markov Chain Monte Carlo (MCMC) described in a previous work
(G13).  The spectral model used is the NS hydrogen atmosphere model
{\tt nsatmos} \citep{heinke06a} modulated by galactic absorption
modeled with {\tt wabs} \citep{morrison83}.  In all runs, convergence
is ensured by visual inspection of the parameters traces, and with the
Geweke test \citep{geweke92}.

The previous work accounted for and quantified the dominant sources of
uncertainties, and presented a detailed description of the analysis
assumptions possibly affecting the results (G13). These assumptions,
also used in the present work, include: (1) isotropic NS surface
thermal emission through (2) an un-magnetized
($B\approxlt\ee{9}\unit{G}$) and (3) H atmosphere of (4) non-rotating
NSs.

Others \citep{lattimer14} have extrapolated the previous
\mr\ constraints from qLMXBs (G13) under an altered assumption of
He-atmosphere -- rather than H atmosphere -- unmagnetized transient
NSs with thermal spectra in quiescence, such as have been predicted
\citep{bildsten04,ivanova08}, obtaining a different \rns\ value.
However, to date, no such objects have been observed
\citep{lattimer14}.  In contrast, a literature search finds at least
several H-companion, unmagnetized, transient LMXBs which have had
thermal emission detected in quiescence (e.g., Cen~X-4, Aql~X-1, 4U
1608-522, MXB 1659-29, XTE~J2123-058).

The present work assumes that all of the qLMXB targets have
unmagnetized H atmospheres.  While it is not observationally
demonstrated that all of these targets are H atmosphere, and not He
atmosphere, one (\OmCen) is observed with $H\alpha$ disk emission,
indicating a H-companion, and so a H atmosphere NS \citep{haggard04}.

Under the forgoing assumptions, the {\tt nsatmos} photon spectral
emission model is fit to data from each qLMXB, with 5 model parameters
for each source ($i$) consisting of: (1) the distance to each source;
(2) an effective surface temperature; (3) an equivalent hydrogen
column density, accounting for line-of-sight absorption between the
observer and the source; (4) a NS mass $M_{{\rm NS}, i}$; and (5) a
normalization for a power-law spectral component (unrelated to the
surface emission) with an assumed photon spectral slope of 1. The
final parameter, $R_{{\rm NS}}$, is alternatively an independent, free
parameter common to all NSs in the \CstR\ model; or is dependent on
$M_{{\rm NS}, i}$ when dEOS models are imposed.

The distances to each NS are included using Gaussian Bayesian prior
distributions derived from previous GC distance measurements: for the
NSs used in the previous work, the same distance priors are used
(G13); for M30, $d_{\rm M30}=9.0\pm0.5\kpc$
\citep{carretta00,lugger07}.

\begin{figure*}
  \centering {~\psfig{file=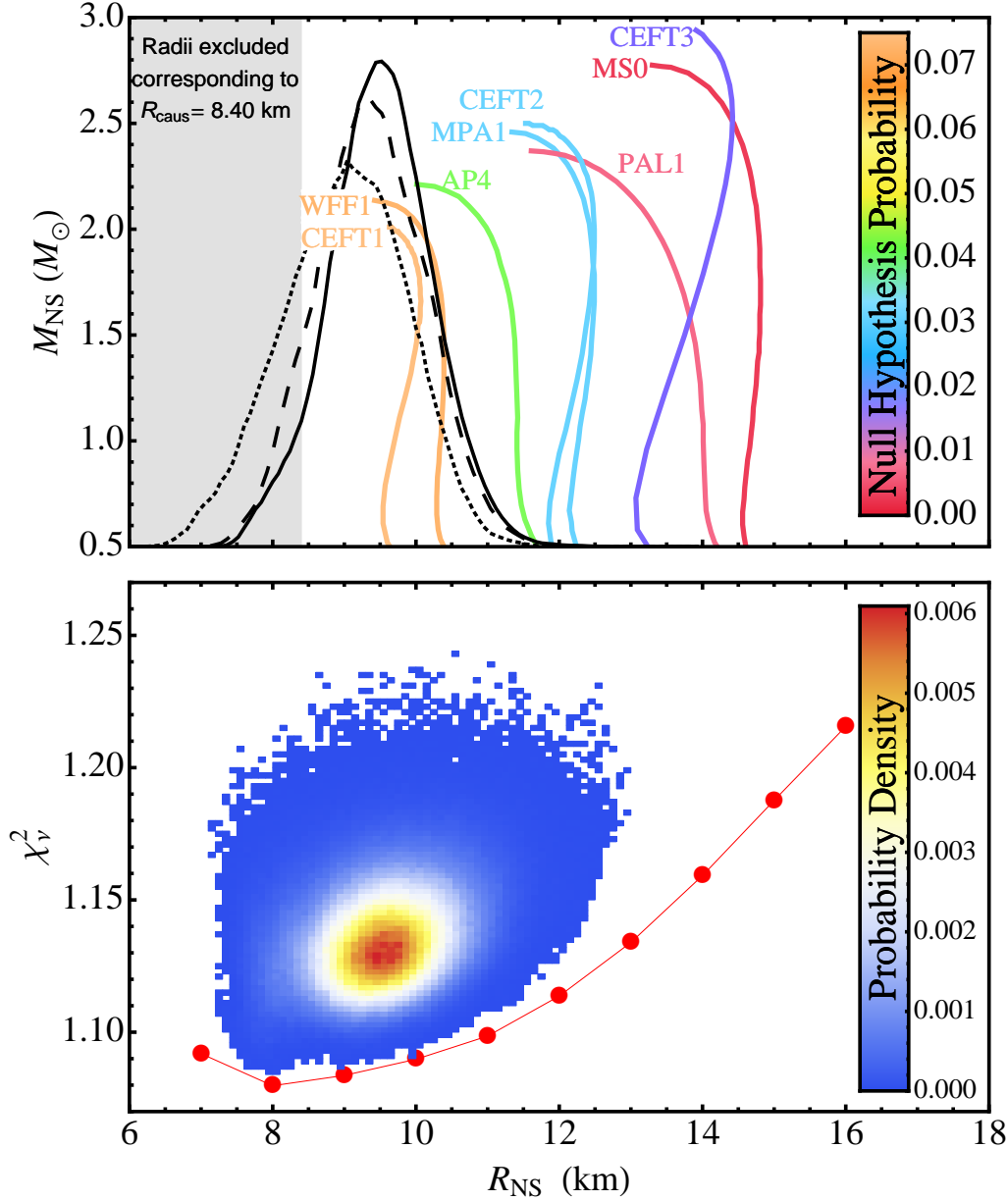,width=15cm,angle=0}~}  
    \caption[]{\small {\it (top)} In black, the normalized posterior
      PDFs of \rns\ are shown, in the three cases of the
      \CstR\ assumption listed in Table~\ref{tab:constantR}:
      considering causality (solid black curve), without considering
      causality (dashed black curve), and the \rns\ PDF of the
      previous work (G13), respectively.  The \mns\ axis does not
      apply to these PDFs. The gray area represents radii excluded by
      $R_{\rm caus}<2.83 G\mns / c^{2}$, resulting from the existence
      of a 2.01\msun\ NS respecting causality.  The tested dEOSs in
      \mr\ space are color-coded by null hypothesis probabilities
      (NHPs, right color bar).  PAL1 and MS0 are rejected with 99\%
      confidence, since their NHPs are smaller than 0.01.  {\it
        (bottom)} Best reduced \chisqnu\ obtained when fixing \rns\ to
      the values in red (from 7\km\ to 17\km), and 2D distribution of
      the \chisqnu\ for accepted radius values in the MCMC run with
      the \CstR\ assumption.}
    \label{fig:results}
\end{figure*}

\section{Constant \rns\ analysis}
\label{sec:cstR}

In this Letter, the same qLMXBs as the previous work are
included in the sample (G13): the qLMXBs in the GCs M28, NGC~6397,
M13, \OmCen, and NGC~6304, observed with the \chandralong\ ACIS-S/I
detectors or with the \xmmlong\ pn camera.  To these, the spectrum of
the qLMXB in M30 observed with \chandra\ (ObsID 2679,
\citealt{lugger07}) and 200\ksec\ of recently archived
\chandra\ observations of the qLMXB in the GC \OmCen\ (ObsIDs 13726
and 13727) are added.  These added spectral data increase the S/N by
7\% compared to the previous analysis (G13).  Data processing and
analysis follow the standard procedures, previously used and described
in detail (G13).  However, \chandra\ and \xmm\ data have been
re-processed with CIAO v4.5 (with CALDB v4.5.5, \citealt{graessle07})
and XMMSAS v12.0.

\begin{figure*}
  \centering {~\psfig{file=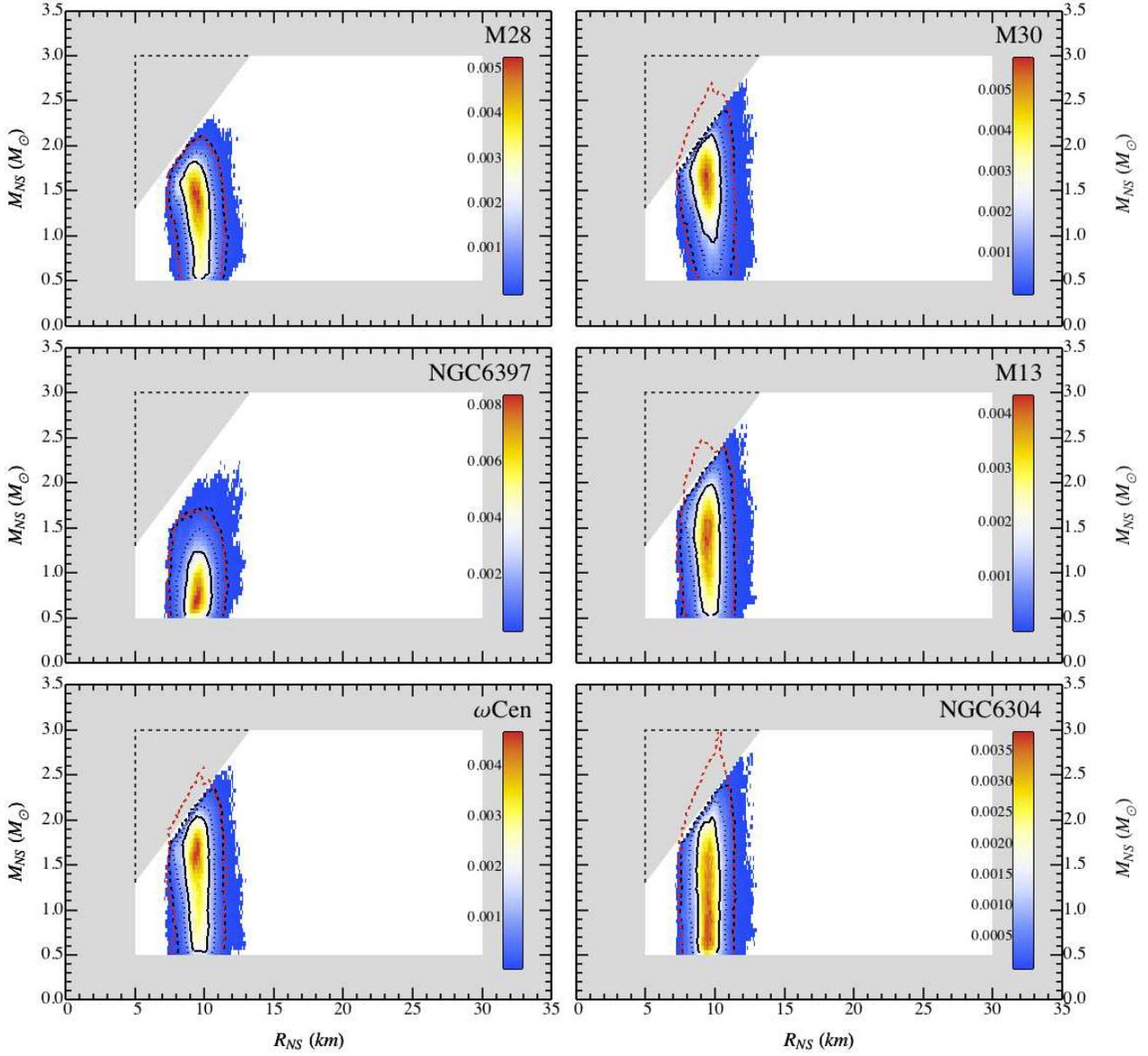,width=17cm,angle=0}~}
    \caption[]{\small Figure showing the posterior distributions in
      \mr\ space for the \CstR\ toy model.  The gray-shaded areas show
      the priors on \mns\ and \rns (hard limits of {\tt nsatmos},
      dashed straight lines), to which the causality-imposed exclusion
      region is added ($\rns < 2.83 GM_{{\rm NS}, i} / c^{2}$).  Black
      solid, Dotted and Dashed contours correspond to the 68\%, 90\%
      and 99\% enclosures.  The red contour is the 99\% enclosure
      obtained without the causality assumption.}
    \label{fig:MR}
\end{figure*}

For \OmCen, the addition of $\sim$200\ksec\ of spectral data improves
the constraints on \rinfty\ and \nh, when analyzed alone.
Specifically, $\rinfty=13.5\ud{4.2}{3.0}\km$ at $d=4.8\kpc$, using all
the available data, compared to $\rinfty=23.6\ud{7.6}{7.1}\km$ at
$d=4.8\kpc$ using the \xmm\ and 2000 \chandra\ data in G13.  A recent
publication \citep{heinke14} also reports this large difference in the
measured radius: $\rns = 20.3\ud{9.1}{6.7}\km$ at 1.4\msun\ with the
same data as G13, and $\rns=11.5\ud{3.4}{3.5}\km$ at 1.4\msun\ when
all available data are used, both assuming $d=5.3\kpc$, and with the
{\tt wabs} absorption model.  Using the $d=5.3\kpc$, we find that the
physical radius at 1.4\msun\ is $\rns=11.7\ud{3.6}{5.3}\km$,
consistent with that of \cite{heinke14}.  Furthermore, the choice of
absorption model does not significantly change the radius
measured. The \rns\ are consistent with the measured \rns\ of
\cite{heinke14} whether the {\tt wabs} or {\tt wilms}
($\rns=10.0\ud{2.6}{5.0}\km$) models are used.

Using these \xray\ data in the MCMC analysis framework, the
\CstR\ analysis of the previous work (G13) is first reproduced.  An
additional analysis is performed for comparison purposes.  It employs
the assumption of causality, imposing that the speed of sound in the
NS core should not exceed the speed of light, i.e., $dP/d\rho \leq
c^2$.  This assumption produces a NS radius $R_{\rm caus} = 2.83 G\mns
/ c^{2}$ above which, the NS would be unstable against collapse
\citep{lattimer07}.  Thus, ``imposing causality'' in this analysis
means that MCMC trials in which $\rns < 2.83 GM_{{\rm NS}, i} / c^{2}$ are
summarily rejected.

It is found that the \rns\ posterior probability distribution function
(PDF) measured with or without the assumption of causality are
consistent with each other and with that of the previous work (G13).
These results are shown in Table~\ref{tab:constantR} and
Figure~\ref{fig:results} (top panel, black lines).  With 90\%
confidence, the resulting NS radius is $\rns=9.4\pm1.2\km$.  When
causality is imposed, $\rns=9.5\pm1.2\km$ \footnote{It should not be
  necessary to impose the causality assumption in this analysis as,
  surely, nature imposes it for us.}.  Figure~ \ref{fig:MR} shows the
posterior distributions in \mr\ space resulting from the
\CstR\ toy model run (without and with the causality assumption,
respectively), and the hard priors imposed in this analysis.

Note, importantly, that imposing the causality constraint does not
shift the \CstR\ posterior PDF to higher radii (Figure~\ref{fig:MR}),
but simply rejects parts of the \mr\ parameter \mr\ space.  The
existence of large-mass NSs \citep{demorest10,antoniadis13} combined
with the causality condition leads to a minimum acceptable value of
\rns.  In other words, the effect of larger mass NSs being detected is
to exclude the part of the \CstR\ posterior PDF below the
corresponding $R_{\rm caus}$.  For example, for $\mns=2.01\msun$,
$R_{\rm caus}=8.40\km$ excludes \rns\ parameter space below this value
(see Figure~\ref{fig:results}).  As a consequence, there is
significant tension between the \CstR\ result and high-mass NSs.
Therefore, a definitive measurement of a 2.6\msun\ NS -- within the
uncertainty of a $2.40\ppm0.12\msun$ pulsar \citep{vankerkwijk11} --
would rule out a \CstR\ radius $<11.0\km$, essentially all the
\rns\ acceptable parameter space obtained here.  This would falsify
the conclusions of this work, and therefore one or more of the
assumptions made.

\begin{deluxetable}{lccccccc}
  \tablecaption{\label{tab:EoSs} Statistical Rejection of Dense Matter Equations of State} 
  \tablewidth{0cm}
  \tabletypesize{\scriptsize}
  \tablecolumns{5}
  \tablehead{
    \colhead{Equation} & \colhead{Reference} & \colhead{\chisq / dof} & \colhead{Null Hypothesis} & \colhead{Acceptance}\\ 
    \colhead{of State} & \colhead{}          & \colhead{}             & \colhead{Probability}     & \colhead{Rate \%}   \\ 
  }
  \startdata 
  WFF1 & \cite{wiringa88} & 523 / 477 & 0.073 & 6.6\% \\ 
  AP4  & \cite{akmal97}   & 531 / 477 & 0.044 & 7.2\% \\ 
  MPA1 & \cite{muther87}  & 536 / 477 & 0.031 & 7.5\% \\ 
  PAL1 & \cite{prakash88} & 557 / 477 & 0.007 & 8.5\% \\ 
  MS0  & \cite{muller96}  & 565 / 477 & 0.003 & 8.2\% \\ 
  \hline                                                 
  \hline                                                 
  CEFT1 & \cite{hebeler13} & 520 / 477 & 0.084 & 6.5\% \\    
  CEFT2 & \cite{hebeler13} & 536 / 477 & 0.031 & 7.3\% \\    
  CEFT3 & \cite{hebeler13} & 547 / 477 & 0.014 & 7.8\% \\
  \enddata

  \tablecomments{The dEOS are ordered by increasing average radius.
    The \chisq-statistics listed is the resulting minimum
    \chisq\ obtained from the MCMC analysis described in the text.}
\end{deluxetable}

\section{Testing proposed dEOSs}
\label{sec:dEOSs}

\begin{figure*}
  \centering {~\psfig{file=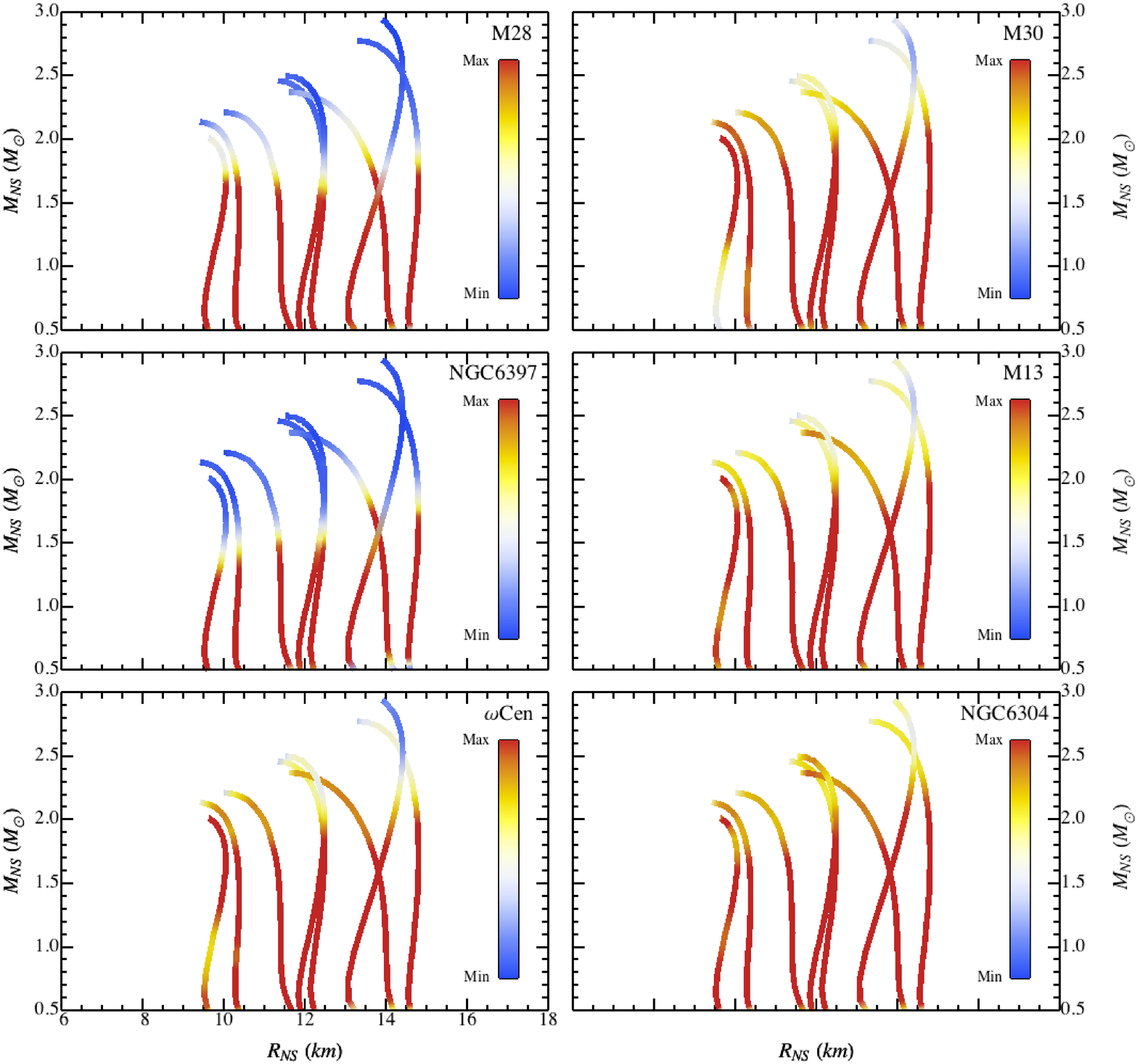,width=17cm,angle=0}~}
    \caption[]{\small Figure showing the posterior distributions in
      \mr\ space for all the tested dEOSs.  Each curve is colored by
      the value of the PDFs (different for each dEOSs, therefore the
      color bar only shows ``Min'' and ``Max'').  In the MCMC runs
      testing dEOS models, the \mns\ and \rns\ are constrained to
      remain on the dEOS with the use of a position parameter sampling
      the \MofR\ curve between 0.5\Msun\ and $M_{\rm max}$.
      Therefore, the priors for \mns\ and \rns\ are the \MofR\ curves
      of the dEOSs.}
    \label{fig:MS0_MR}
\end{figure*}

Here, it is measured to what degree of certainty a selection of
popular proposed dEOSs fit the spectra of the six qLMXBs in the
sample.  This spectral analysis is performed with the MCMC analysis
described above by forcing the fitted \mns\ and \rns\ parameters to be
constrained to the \MofR\ curve of a user-defined dEOS.  To do so, the
MCMC-sampled parameters are not the \mns\ and \rns\ of each NSs, but
instead a parameter between 0 and 1 representing the position on the
dEOS curve, with 0 at the tip (at $M_{\rm max}$), and 1 at the
$\mns=0.5\msun$ point (see posterior PDFs in Figure~\ref{fig:MS0_MR}).

To determine which dEOS is quantitatively preferred by the data, five
dEOSs with varied stiffness are tested, as well as three
representative dEOSs obtained from chiral effective field theory
\citep{hebeler13}: the Soft, Intermediate and Stiff representative
dEOSs (labeled CEFT1, CEFT2, and CEFT3 here).  For each dEOS, the
minimum \chisq\ and the corresponding null hypothesis probability
(NHP) are obtained after convergence of the MCMC simulation.  The NHP
is the probability of finding by chance a \chisq\ as large or larger
as the minimum \chisq\ found if the model is correct.  Results are
shown in Table~\ref{tab:EoSs} and Figure~\ref{fig:results}.  The MCMC
runs testing the eight dEOSs have acceptance rates in the range
$\sim6-9\%$.  In addition, one can notice that the NHP decreases as
the average radius, i.e. the stiffness, of the tested dEOS, increases.

To the reader, there may appear to be a discrepancy between the NHPs
obtained when testing dEOSs and the likelihood distributions of the
\CstR\ case (Figure~\ref{fig:results}, {\it top}).  Specifically, on
the one hand, the posterior PDFs of \rns\ indicates that $\rns=14\km$
is $\sim6\sigma$ away from the median of the distribution.  On the
other hand, the test for MSO (with average radius $\sim14\km$) gives a
NHP of 0.003, i.e., MS0 is rejected at the 99.7\% confidence
($\sim$3$\sigma$).

This is better illustrated with Figure~\ref{fig:results} ({\it
  bottom}) showing the 2D distribution of the \chisqnu\ value as a
function of \rns\ for all accepted MCMC steps of the \CstR\ analysis
(no causality assumed), revealing the extent of the radius coverage.
The plot also shows the minimum \chisqnu\ (red) obtained when the
\rns\ is fixed to a selection of values (7--16\km), essentially a
selection of toy model dEOSs.  The minimum \chisqnu\ of the 2D
distribution follows the minimum \chisqnu\ of the red line in the
range 7--11\km, but becomes systematically larger above 12\km\ and
disappears above 13\km.  This apparent difference (between the
Bayesian-accepted parameter space and the minimum \chisqnu\ values at
each \rns) is similar to the difference between confidence levels
described in the previous paragraph.  This emerges as a consequence of
the complicated parameter space above $12\km$ in the \CstR\ run; the
probability of excursion to radii larger than 12\km\ during the MCMC
run is low since such steps are more likely to be rejected, making it
difficult for the MCMC to populate this area of the parameter space.

This justification is investigated with the rejected MCMC steps in the
\CstR\ analysis.  As the \rns\ goes to larger values, the NS masses
are forced to smaller values to maintain the \rinfty\ values that fits
the spectral data of each NS.  A constant $\rinfty = \rns \left ( 1-
2G\mns / \rns\,c^{2} \right)^{-1/2}$ implies for large \rns, smaller
\mns.  As a result, an increasing proportion of MCMC steps are
rejected because the mass parameters are more likely to lie outside
the allowed range of values (0.5--3.0\msun, the hard limits of the
{\tt nsatmos} model).  A significant fraction (93.2\%) of the proposed
steps with $\rns>13\km$ are rejected simply due to one or more
$M_{{\rm NS}, i}$ being outside the prior limits.  For comparison,
when all radii are considered, only 40.4\% of proposed steps are
rejected due to one or more $M_{{\rm NS}, i}$ being outside the hard
limits.  Interestingly, we note that a larger minimum \mns\ value
(assumed from theory) would result in an even lower maximum radius
from this Bayesian analysis.

To further test that the observation illustrated in
Figure~\ref{fig:results} ({\it bottom}) is a consequence of the shape
of the parameter space, a separate \CstR\ trial is performed, in which
all MCMC walkers were initiated with $\rns>13\km$.  The walkers are
all observed to ``move'' and converge into the final parameter space of
Figure~\ref{fig:results} ({\it bottom}), with all $\rns<13\km$, as
expected from the MCMC algorithm.

The ``noisy'' parameter space occurs because of the strong covariance
between the 32 model parameters in the \CstR\ analysis.  Overall, the
apparent difference between the acceptable parameter space from the
MCMC analysis and the \chisq\ analysis is due to the fact that
Bayesian and \chisq\ statistics rely on different assumptions about
the statistical system.  This underscores the need for a Monte Carlo
integration to assess most-likely parameters and uncertainty regions,
rather than relying upon the \chisq\ curvature matrix.

Overall, this minimum \chisq\ values obtained for each dEOS tested in
this analysis, and their corresponding NHPs, provide a quantitative
confrontation between the proposed physical models and the
\xray\ spectral data of qLMXBs.

\section{Conclusion}
\label{sec:ccl}

This work first updated the \CstR\ measurement presented in a previous
work (G13), including additional spectral data (for the qLMXBs in
\OmCen\ and M13), and finding $\rns=9.4\pm1.2\km$ (90\% confidence),
consistent with the previous work.  It was found that the inclusion of
causality does not significantly affect the posterior PDFs in the
\CstR\ toy model.

In the second part of this work, the simultaneous spectral analysis
was performed by constraining the \mns\ and \rns\ parameters of the
six qLMXBs on the \MofR\ relations of a selection of tested EOSs.  The
resulting statistics from the MCMC spectral fitting permits rejection
of the dEOSs PAL1 and MS0 on the basis of the NHPs obtained, 0.007 and
0.003, respectively.

In conclusion, two of the dEOSs tested (PAL1, \citealt{prakash88} and
MS0, \citealt{muller96}) have ${\rm NHP}<0.01$, rejecting these
theoretical dEOSs as adequate descriptions of the behavior of cold
nuclear matter with $>$99\% confidence, under the assumptions of this
work.  This is the first time that a selection of proposed dEOSs are
conservatively excluded with quantitative probabilities of
consistency, using an analysis of qLMXB \xray\ Spectra.

{\bf Acknowledgements} S.G. and R.E.R acknowledge the support of NSERC
via the Vanier CGS and Discovery grant programs. The authors
acknowledge useful and enlightening conversation with David W. Hogg on
MCMC particulars.

\bibliographystyle{apj_8}

\end{document}